\documentstyle[11pt,newpasp,twoside,epsfig]{article}
\def\arg#1{{\it#1\/}}

\def\edcomment#1{\iffalse\marginpar{\raggedright\sl#1\/}\else\relax\fi}
\reversemarginpar

\begin{document}
\title{Extended X-ray emission from FRIIs and RL quasars}
\author{G. Setti, G. Brunetti}
\affil{Dipartimento di Astronomia, Universit\'a di Bologna}
\affil{Istituto di Radioastronomia del CNR, Bologna}
\author{A. Comastri}
\affil{Osservatorio Astronomico di Bologna}

\begin{abstract}

We review the evidence that detectable fluxes of X-rays are produced
by inverse Compton scattering of nuclear photons with the 
relativistic electrons in the radio lobes of strong FRII 
radio galaxies within the FRII-RL 
quasar unification scheme. We report here on the possible detection of
this effect in two steep spectrum RL quasars. 
This may have important implications on the 
physics and evolution of powerful radio galaxies.

\end{abstract}

\section{Introduction}

It is well known that the synchrotron radio emission from the 
extended lobes of strong radio galaxies and radio-loud (RL) quasars
samples ultra relativistic electrons. It is a customary practice
to estimate the average magnetic field intensity via the minimum
energy argument by making use of the emitted radio flux in the
10 MHz -- 100 GHz band (source rest frame). Since the critical 
frequency emitted in the synchrotron process is 
$\nu(MHz) \sim B(\mu G) (\gamma/10^3)^2$ 
and typically $B(\mu G) > 10$, it then follows that
only electrons with Lorentz factor $\gamma > 10^3$ are taken into
account. These electrons are also responsible for producing X-rays
via the inverse Compton (IC) scattering of the cosmic 
microwave background (CMB) photons 
[$\epsilon(keV) \sim (\gamma/10^3)^2$]. 
When these X-rays are detected, 
the number of ultra-relativistic electrons is fixed and from 
the radio flux one can uniquely determine the average
magnetic field intensity. Up to now, because of the weakness and diffuse
nature of the 
predicted X-ray emission, detection of X-ray fluxes due to this
process has been possible for a few sources only, notably Fornax A
(Feigelson et al. 1995; Kaneda et al. 1995) and 
Cen B (Tashiro et al. 1998); 
the derived magnetic field intensities are lower than 
the classical equipartition values by factors 1.5--2.

Brunetti, Setti \& Comastri (1997) have pointed out that 
sizeable X-ray fluxes
can also be emitted by the IC scattering of relativistic electrons 
in the FRII's radio lobes with the IR photons from a quasar, 
and associated circumnuclear 
dusty/molecular torus, hidden in the galaxy's nucleus. 
Since the IR emission peaks at $50-100 \mu m$ 
electrons at lower energies ($\gamma < 500$) 
are involved in this process. Of course there
are no reasons why these lower energy particles shouldn't be present,
on the contrary one would expect them on physical grounds based on 
acceleration and loss mechanisms. In order to estimate the size
of the expected X-ray flux, for a given quasar IR emission,
one may work out the equipartition by extrapolating downward 
the electron spectrum derived from the synchrotron emission to a
minimum energy ($\gamma_{min}$) limited from below by possible Coulomb 
losses. The equipartition fields ($B_{eq}$) so derived are stronger than
the classical one by factors from 1.5 to 3 
(also Setti, Brunetti \& Comastri 1999). 
We have 
shown that the IC scattering of the IR nuclear photons may easily 
account for a large fraction of the extended X-ray emission of several 
powerful FRIIs
at large redshifts ($z \sim 1$) detected by ROSAT in the 0.1--2.4 keV 
interval. 

Morphologically there are two important aspects that should be 
mentioned: firstly, for obvious geometrical reasons, the X-rays 
from the IC scattering of the quasar photons
are more concentrated toward the nuclear region 
than those from the IC scattering of the 
CMB photons and, secondly, given two symmetrical radio lobes the 
X-ray emission from the far lobe can be much larger than that 
from the near one, depending on the orientation of the radio
axis with respect to the line of sight, due to the enhanced 
efficiency of head-on scatterings (Brunetti et al.1997). 
It should also be mentioned that, while the X-rays from the IC 
scattering of the CMB photons must have a spectral slope coincident 
with that of the synchrotron radio emission, the X-ray spectrum 
associated with 
the IC scattering of the nuclear photons may or may not have
the synchrotron slope simply because a different 
portion of the primary electron spectrum is being sampled
(see also Brunetti 2000). 

Direct evidence of extended X-ray emission from the IC scattering 
of the IR photons from a hidden quasar has been gathered by 
Brunetti et al.(1999) making use of ROSAT HRI observations of 
the powerful, double lobed radio galaxy 3C 219. The residual 
X-ray distribution after subtraction of the absorbed, unresolved 
nuclear source is remarkably coincident with the radio structure.
The central extended ($\sim$ 100 kpc) X-ray emission, somewhat stronger
in the counter-jet side as expected in our model, can be accounted for
by assuming a magnetic field 
$\sim 3$ times weaker than our equipartion value ($B_{eq} 
\simeq 10\mu G$, 
$\gamma_{min} = 50$). Of course this estimate depends on the assumed
IR power of the hidden quasar which we have derived by two, albeit
indirect, approaches since 3C 219 has not been detected by IRAS 
and not observed by ISO. Observations with {\it Chandra}, 
scheduled in the fall of the year 2000, will likely provide a check 
of our model.


\begin{figure*}
\centerline{
\psfig{figure=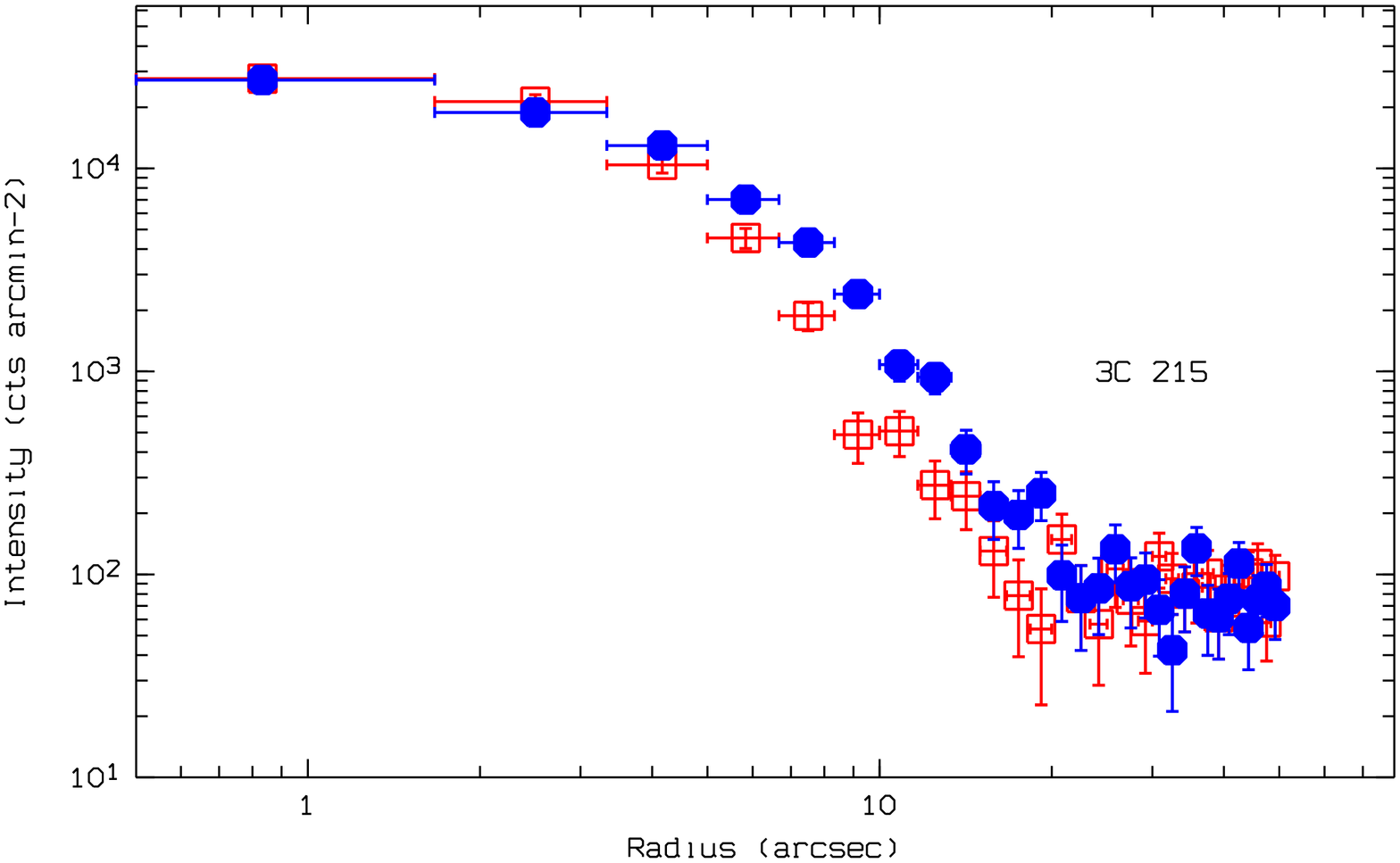,width=11cm,angle=270}
}
\end{figure*}

\section{IC X-rays from the radio quasars 3C 215 and 3C 334} 

Extended X--ray emission around RL quasars 
has been recently discovered from the analysis 
of ROSAT HRI observations
(Crawford et al. 1999; Hardcastle \& Worrall 1999).
The origin of the extended and rather weak component (5--15\% of total 
intensity) is usually 
ascribed to thermal emission from the surrounding
intracluster medium (ICM), although a significant contribution
from the IC scattering of the quasar's photons predicted
by our model cannot be excluded (Crawford et al. 1999).

In order to test the IC hypothesis 
we have carried out a detailed analysis of the spatial
profiles of the sources in the Crawford et al.(1999) sample.
The data retrieved from the public archive have been analyzed 
following a procedure similar to that in Crawford et al.(1999),
but the azimuthal distribution has been investigated with the aim
of checking for a possible correlation with the radio structure. 
Accordingly
the source counts have been subdivided in four quadrants centered
on the quasar (X-ray source peak) and so oriented that two opposite
quadrants are aligned with the radio axis defined by the direction
of the innermost radio lobes. We have then compared the source counts
in the two quadrants along the radio axis with those collected in the
perpendicular direction. 



We find evidence of X--ray extension spatially 
correlated with the radio axis in two  
quasars: 3C 215 and 3C 334. For 3C 48, 254 and 273 our analysis is
inconclusive since their radio angular sizes are lower then, or 
comparable to, the HRI PSF. For 3C 215 (Fig.1) 
the effect is strong : a KS test rejects at 99.9\% level 
the hypotesis that 
the count distributions along (filled dots) and
perpendicular to (open squares) the radio axis are extracted from
the same population. In the case of 3C 334 (not shown here),
although the count distribution along the radio axis systematically
exceeds that in the perpendicular direction, the effect is
statistically marginal (the same KS test gives $\sim 92\%$).

In order to check whether the elongation on $\sim 10$ arcsec
scale could be due to an intrinsic elongation of the
HRI PSF and/or to an insufficient correction of the  
wobbling, we have analyzed several isolated stars (companion 
at least 6 mag fainter) 
with similar count statistics extracted
from the RASSDWARF catalogue (Huensch et al. 1998): no evidence
of asymmetric distributions has been found. Moreover the count profiles
of 3C 215 and 334 in the direction perpendicular to the radio axis
are consistent with those of spatially unresolved sources.

The X--ray fluxes (0.1--2.4 keV, spectral index $\alpha = 1$) 
associated with the extended structures
have been estimated by   
subtracting the counts within the 
quadrants in the direction perpendicular to the radio axis
from those within the quadrants aligned with radio axis: the
luminosity of 3C 215 is $\sim 1.5 \cdot 10^{45}$ erg s$^{-1}$ of which
$\sim 2.2 \cdot 10^{44}$ erg s$^{-1}$ in the extended component,
while for 3C 334 one has $\sim 10^{45}$ erg s$^{-1}$
and $\sim 10^{44}$ erg s$^{-1}$, respectively 
[$H_o = 75$ km/s/Mpc, $q_o = 0$].

Knowledge of the quasar IR radiation is of crucial importance for the
computation of the expected IC X-ray fluxes. Unfortunately no FIR data 
are available 
for 3C 215, while 3C 334 has been observed by IRAS with a 
60$\mu m$ luminosity of $\sim 10^{46}$erg s$^{-1}$
(van Bemmel et al. 1998). By adopting typical quasar SEDs 
we estimate a 
$1 - 100 \mu m$ luminosity of $\sim 10^{46}$ erg s$^{-1}$  
and $\sim 4 \cdot 10^{46}$ erg s$^{-1}$ for 3C 215 and 3C 334,
respectively. Following Brunetti et al.(1997) model
we find that the magnetic field intensities 
required to fully account for the X-ray fluxes of the extended 
components are $\sim 5$ and $\sim 3$ times smaller than $B_{eq}$ 
for 3C 215 and 334, respectively. In each source $B_{eq}$ has been 
calculated by extrapolating downward to $\gamma_{min} = 50$ 
the power law electron spectrum derived from the low frequency 
radio spectrum. It should be pointed out that by applying the 
standard
equipartition we would have obtained factors $\sim 2.6$ (3C 215) and
$\sim 2$ (3C 334) below the corresponding equipartion fields, but this
would be conceptually wrong.

\section{Conclusions}

There is supportive observational evidence of 
extended X-ray emission from the IC scattering of quasar 
IR photons with relativistic electrons in the lobes 
of powerful radio galaxies. Besides being a 
confirmation of the FRII--quasar unification, this may provide an 
important tool for the diagnostic of the relativistic plasma 
at particle energies not sampled by radio observations. 
The unavoidable presence of lower energy particles implies 
stronger magnetic fields than derived by standard equipartition 
formulae and, consequently, a larger pressure inside the lobes. 
Moreover, accounting for the extended X-ray emission in sources 
for which the quasar radiation can be constrained indicates 
magnetic field strengths lower than equipartition. 
Therefore, confirmation of our IC model by {\it Chandra} and 
XMM satellites may 
provide important clues on the physics and evolution of radio 
sources.  

\acknowledgments
It is a pleasure to thank C.S. Crawford, 
A.C. Fabian and I. Lehmann for informative discussions 
concerning 3C 215 and 3C 334.

\end{document}